\documentclass[twocolumn,showpacs,preprintnumbers,prl,amsmath,amssymb]{revtex4}
\usepackage{graphicx}% Include figure files
\usepackage{dcolumn}% Align table columns on decimal point
\usepackage{bm}% bold math

\begin{document}
%\preprint{DV/PDAC}

\title{Phase transitions and iron-ordered moment form factor in LaFeAsO}

\author{H.-F. Li$^1$\footnote{Email: hfli@ameslab.gov.}, W. Tian$^1$, J.-Q. Yan$^1$, J. L. Zarestky$^1$, R. W. McCallum$^{1,2}$, T. A. Lograsso$^{1}$,
and D. Vaknin$^{1,3}$\footnote{Email: vaknin@ameslab.gov.}}

\affiliation{
$^1$Ames Laboratory, US-DOE, Ames, Iowa 50011, USA \\
$^2$Department of Materials Science and Engineering, Iowa State University, Ames, Iowa 50011, USA \\
$^3$Department of Physics and Astronomy, Iowa State University, Ames, Iowa 50011, USA}

\date{\today}
\begin{abstract}

Elastic neutron scattering studies of an optimized LaFeAsO single crystal reveal that upon cooling, an onset of the tetragonal (T)-to-orthorhombic (O)
structural transition occurs at $T_\texttt{S} \approx 156$ K, and it exhibits a sharp transition at $T_\texttt{P} \approx 148$ K. We argue that in the
temperature range $T_\texttt{S}$ to $T_\texttt{P}$, T and O structures may dynamically coexist possibly due to nematic spin correlations recently proposed
for the iron pnictides, and we attribute $T_\texttt{P}$ to the formation of long-range O domains from the finite local precursors. The antiferromagnetic
structure emerges at $T_\texttt{N} \approx 140$ K, with the iron moment direction along the O \emph{a} axis. We extract the iron magnetic form factor and
use the tabulated $\langle j_0\rangle$ of Fe, Fe$^{2+}$ and Fe$^{3+}$ to obtain a magnetic moment size of $\sim$0.8 $\mu_\texttt{B}$ at 9.5 K.

\end{abstract}

\maketitle

\section{I. INTRODUCTION}

The discovery of \emph{R}FeAs(O$_{1-\texttt{x}}$F$_\texttt{x}$) (\emph{R} = rare earth, $\texttt{"}$1111$\texttt{"}$) superconductors \cite{Kamihara2008}
with transition temperatures up to 56 K \cite{Chen2008, Ren2008} has stimulated a renewed excitement in the search for novel superconductors derived from
antiferromagnetic (AFM) parent compounds. In these iron-arsenide based compounds, the superconducting (SC) state can be achieved by doping or by the
application of pressure. The appearance of superconductivity is normally accompanied by a suppression of the AFM state in the parent compounds and by the
appearance of a spin resonance \cite{Christianson2008-1,Lumsden2009-1}. It is thus important to understand the magnetism in these iron-based parent
compounds in order to unravel the mechanism that leads to superconductivity.
\begin{figure}
\centering
\includegraphics[width = 0.46\textwidth] {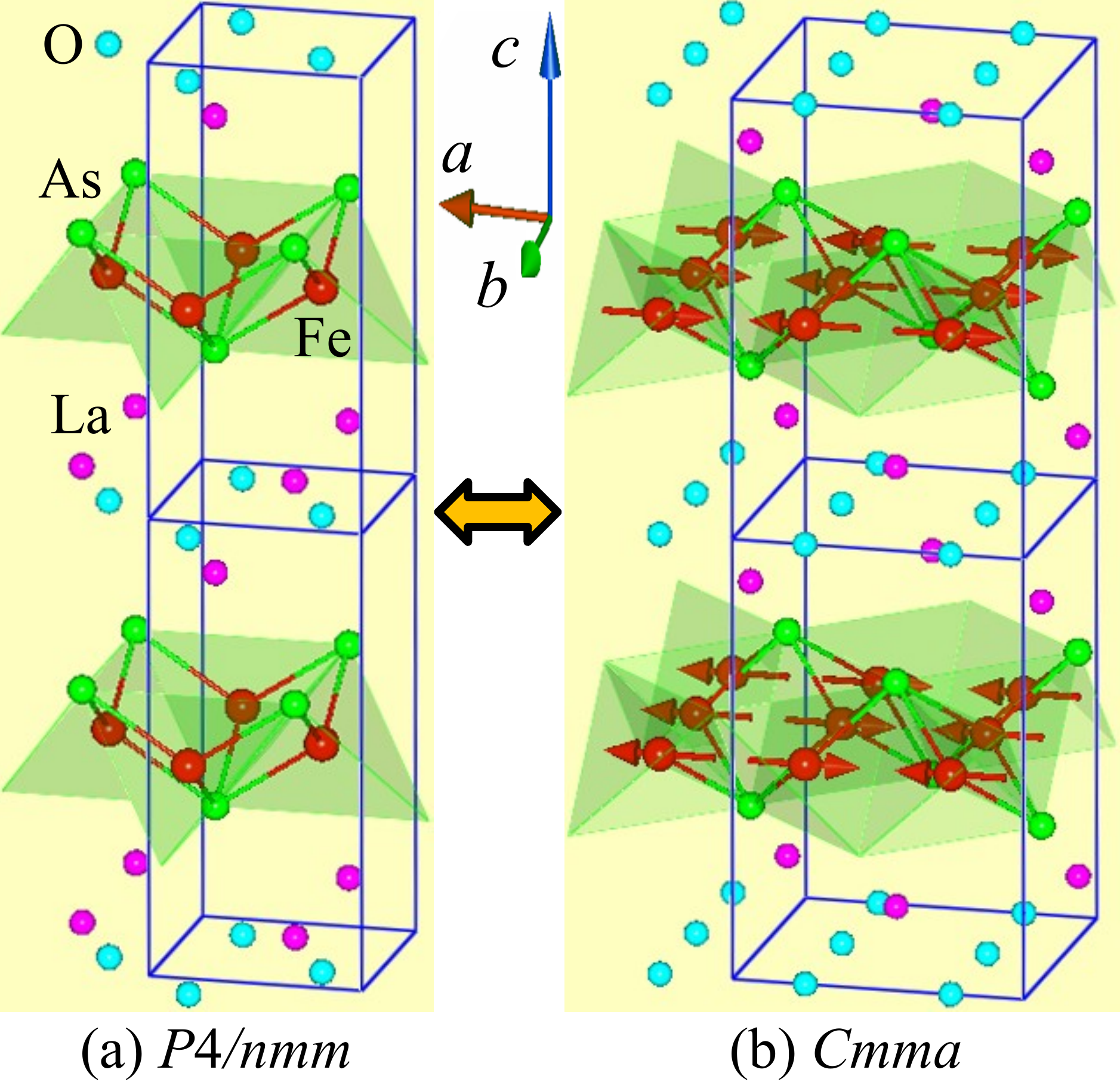}
\caption{(color online) (a) Crystal structure ($P4/nmm$) above $T_\texttt{S}$ = 156(1) K with two unit cells (solid lines). (b) Crystal structure
($Cmma$) below $T_\texttt{S}$ with two unit cells (solid lines) and AFM structure below $T_\texttt{N}$ = 140(1) K in one AFM unit cell.
The arrows on the Fe irons in (b) represent the spins of iron in single-crystal LaFeAsO. The unit cells of $P4/nmm$ (T),
$Cmma$ (O) and AFM structures are (\emph{a b c}) (\emph{a} = \emph{b}), ($\sqrt{2}a$ $\sqrt{2}b$ \emph{c}) and ($\sqrt{2}a$ $\sqrt{2}b$ 2\emph{c}),
respectively.}
\label{structures}
\end{figure}
\begin{figure}
\centering
\includegraphics[width = 0.46\textwidth] {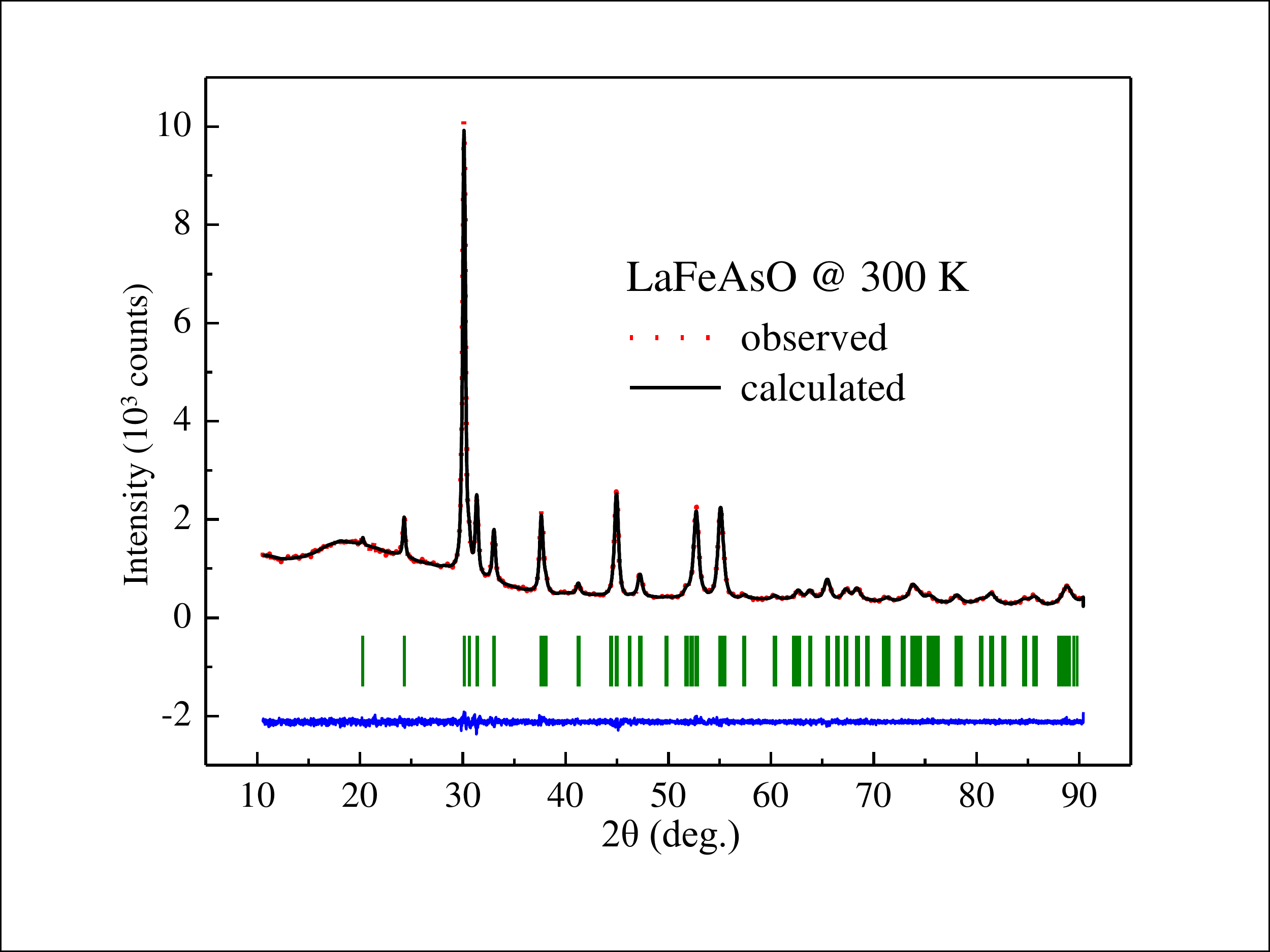}
\caption{(color online) Observed (dots) and calculated (solid line) X-ray powder diffraction patterns
for pulverized LaFeAsO single crystals at ambient conditions obtained on an in-house diffractometer
employing the copper $K_{\alpha1}$ = 1.54056 {\AA} and $K_{\alpha2}$ = 1.54439 {\AA} with $I_{\alpha2}$/$I_{\alpha1}$ = 0.5 as the radiation.
The vertical bars mark the positions of Bragg reflections. The lower curve represents the difference between the observed and calculated patterns.}
\label{powder-diff}
\end{figure}

Because of the initial difficulties to grow large $\texttt{"}$1111$\texttt{"}$ single crystals, most of the research on these systems has been performed on
polycrystalline samples. These studies showed that LaFeAsO undergoes a structural phase transition from the tetragonal (T, $P4/nmm$) symmetry [Fig.\
\ref{structures}(a)] to the orthorhombic (O, $Cmma$) one [Fig.\ \ref{structures}(b)] at $T_\texttt{S}$ $\approx$ 155 K and forms an AFM ordering (also
referred to as a spin-density-wave - SDW) at $T_\texttt{N}$ $\approx$ 137 K upon cooling, with an ordered magnetic moment of 0.36(5) $\mu_\texttt{B}$
\cite{Cruz2008}. Implied in these reports on powdered polycrystalline samples \cite{Cruz2008, Nomura2008, Ishikado2009} is that the form factor used is
that of Fe$^{2+}$ ion. However, first-principle calculations on LaFeAsO predict a larger localized magnetic moment of $\sim$2.6 $\mu_\texttt{B}$ at
each iron site that is embedded in an itinerant electronic environment \cite{Ma2009}. It should be noted that density functional theory argues that a large
enough magnetic moment $\sim$2 $\mu_\texttt{B}$ is necessary to drive the observed O-T transition \cite{Ishibashi2008}. It is clear that there is a
significant discrepancy between the calculated iron moment and the experimental one. Therefore, determining the moment size is vital to the validity of any
theoretical model that attempts to explain the electronic structure of LaFeAsO. Another important issue is the coupling between structural and magnetic
behaviors in iron pnictides. Inelastic neutron scattering studies from polycrstalline LaFeAsO showed two-dimensional magnetic fluctuations that persist up
to room temperature (over $\sim$160 K above $T_\texttt{N}$). It was argued \cite{Li2009-1,Loudon2010,Qureshi2010} that such fluctuations introduce dynamic
disorder of the O/T mixed-phase in the so-called T phase, suggesting that a finite orthorhombicity may exist above the O-T structural transition
\cite{Li2009-1,Qureshi2010, Loudon2010}.
\begin{table}[!ht]
\caption{Refined structural parameters with T ($P4/nmm$) symmetry for pulverized LaFeAsO single crystals at room temperature.\\
$a$ = 4.0316(1) {\AA}, $c$ = 8.7541(1) {\AA}, $V$ = 142.290(1) {\AA}$^3$\\
$R_\texttt{B}$ = 1.01, $R_\texttt{F}$ = 0.60, $\chi^2$ = 1.18} \label{table0}
\begin{ruledtabular}
\begin{tabular} {llllll}
atom & site & x & y & z & $B$ ({\AA}$^2$) \\
\hline
La & 2c   & 0.25 &  0.25 & 0.1405(2)  & 1.65(4)  \\
Fe & 2b   & 0.75 &  0.25 & 0.5        & 1.69(12) \\
As & 2c   & 0.25 &  0.25 & 0.6543(3)  & 2.15(8)  \\
O  & 2a   & 0.75 &  0.25 & 0.0        & insensitive \\
\end{tabular}
\end{ruledtabular}
\end{table}

The polycrystalline studies have provided important insights into the behavior of LaFeAsO, but questions remain with regard to the AFM moment direction,
the moment size and its spatial distribution. To a large extent these questions are related to the complexity of iron chemistry, namely its valence (or
mixed-valence), bonding, and electron configuration in different local chemical surroundings. To address these questions, the study of high-quality single
crystals is vital. Such crystals are now available with the recent successful growth of relatively large $\texttt{"}$1111$\texttt{"}$ single crystals at
ambient pressure \cite{Yan2009}.

Here we report elastic neutron scattering and synchrotron X-ray powder diffraction studies on the bulk and pulverized LaFeAsO single crystals,
respectively, focusing on the details of the structural and magnetic transitions as well as the coupling between them, and the measurement of the average
ordered magnetic moment size and its spatial distribution by extracting the magnetic form factor of iron in this compound.

\section{II. EXPERIMENTAL}

LaFeAsO single crystals were synthesized in an NaAs flux at ambient pressure as described in a recent report \cite{Yan2009}. Crystal quality was
characterized by Laue back-scattering, X-ray powder diffraction, heat capacity, magnetization and resistivity measurements. A large LaFeAsO single crystal
($\sim$20 mg) was selected for this study. The mosaic of this single crystal is 0.59(2)$^\texttt{o}$ full width at half maximum (FWHM) for the
(202)$_\texttt{O}$ (O notation) reflection at room temperature. The elastic neutron scattering measurements were carried out on the HB-1A
fixed-incident-energy (14.6 meV) triple-axis spectrometer using a double pyrolytic graphite (PG) monochromator (located at the High Flux Isotope Reactor,
HFIR, at the Oak Ridge National Laboratory, USA). Two highly oriented PG filters, one after each monochromator, were used to reduce the $\lambda$/2
contamination. The beam collimation throughout the experiment was kept at 48$'$-48$'$-sample-60$'$-360$'$. The single crystal was wrapped in Al foil and
sealed in a He-filled Al can which was then loaded on the cold tip of a closed cycle refrigerator with (\emph{h}0\emph{l})$_\texttt{O}$ in the \emph{Cmma}
symmetry as the scattering plane. The synchrotron X-ray powder diffraction study of pulverized LaFeAsO single crystals from the same batch as the one used
for the neutron diffraction study was carried out at the 11-BM beamline, Advanced Photon Source, Argonne National Laboratory. 11-BM is a bending magnet
beamline, equipped with a vertical beam collimation mirror, a double crystal monochromator with a horizontal sagittal focusing second crystal, and a
vertical focusing mirror. The calibrated X-ray wavelength was 0.41219(1) {\AA}. We note that the (\emph{HKL})$_\texttt{T}$ indices for T symmetry
correspond to the O reflection (\emph{hkl})$_\texttt{O}$ based on the relations $h = H + K, k = H - K$, and $l = L$.

\section{III. RESULTS AND DISCUSSION}
\subsection{A. Structural and magnetic transitions}
\begin{figure}
\centering
\includegraphics[width = 0.46\textwidth] {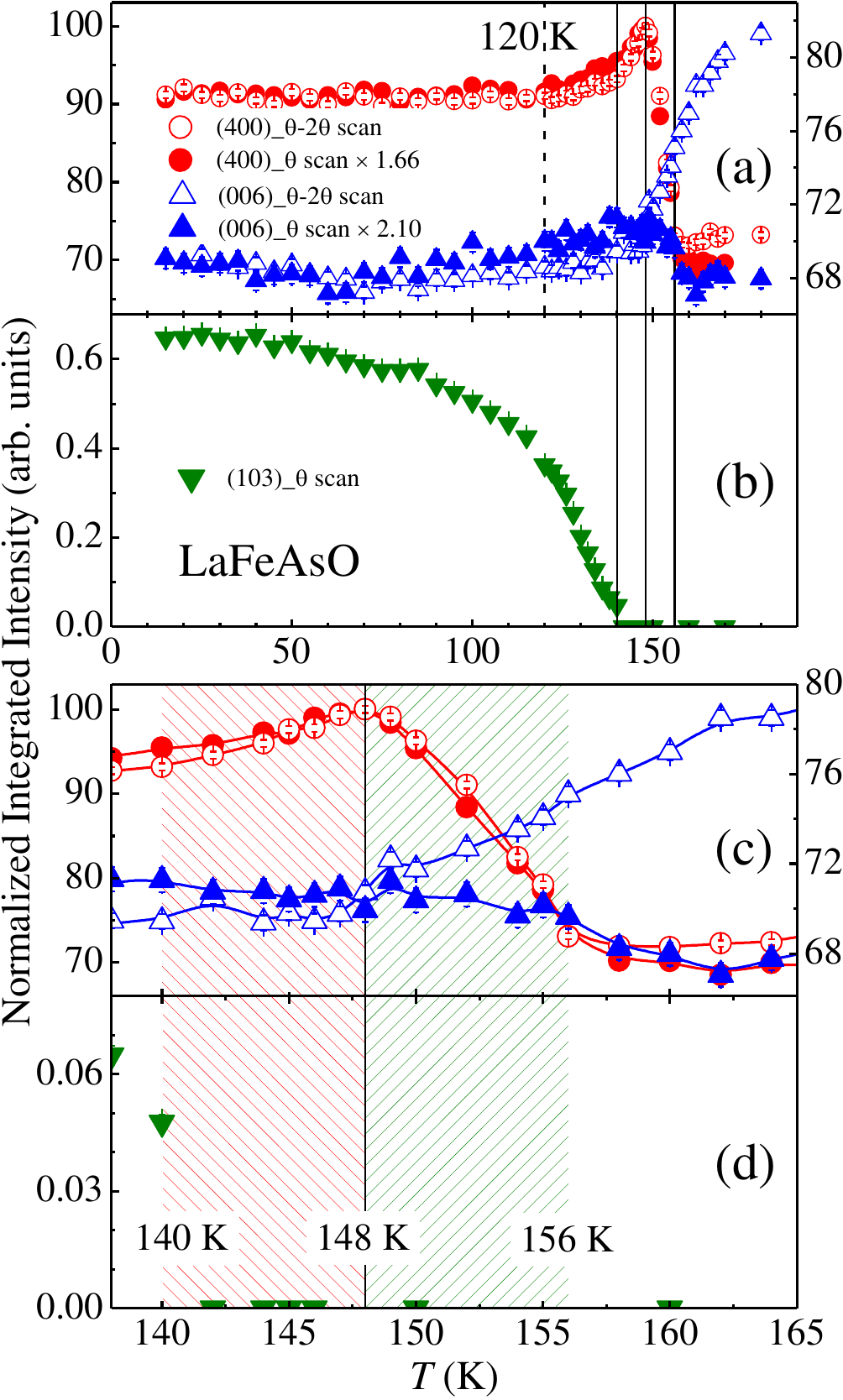}
\caption{(color online)  Temperature dependence of the normalized integrated intensities of (a) $\theta$-2$\theta$ (void symbols) and rocking-curve $\theta$
(solid symbols) neutron diffraction scans of the nuclear Bragg (400)$_\texttt{O}$ (circles) and (006)$_\texttt{O}$ (triangles) reflections, and (b) the
rocking-curve scan of the AFM (103) neutron diffraction reflection (down-solid-triangles) of single-crystal LaFeAsO measured upon warming up the crystal from 9.5 K.
(c) and (d) are the enlargements of (a) and (b) near the structural and AFM transitions, respectively. The integrated
intensities of (400)$_\texttt{O}$ and (006)$_\texttt{O}$ rocking curves are rescaled  by 1.66 and 2.10, respectively, as indicated to make them coincide with
the respective $\theta$-2$\theta$ scans at low temperatures.}
\label{order-params}
\end{figure}
\begin{figure}
\centering
\includegraphics[width = 0.46\textwidth] {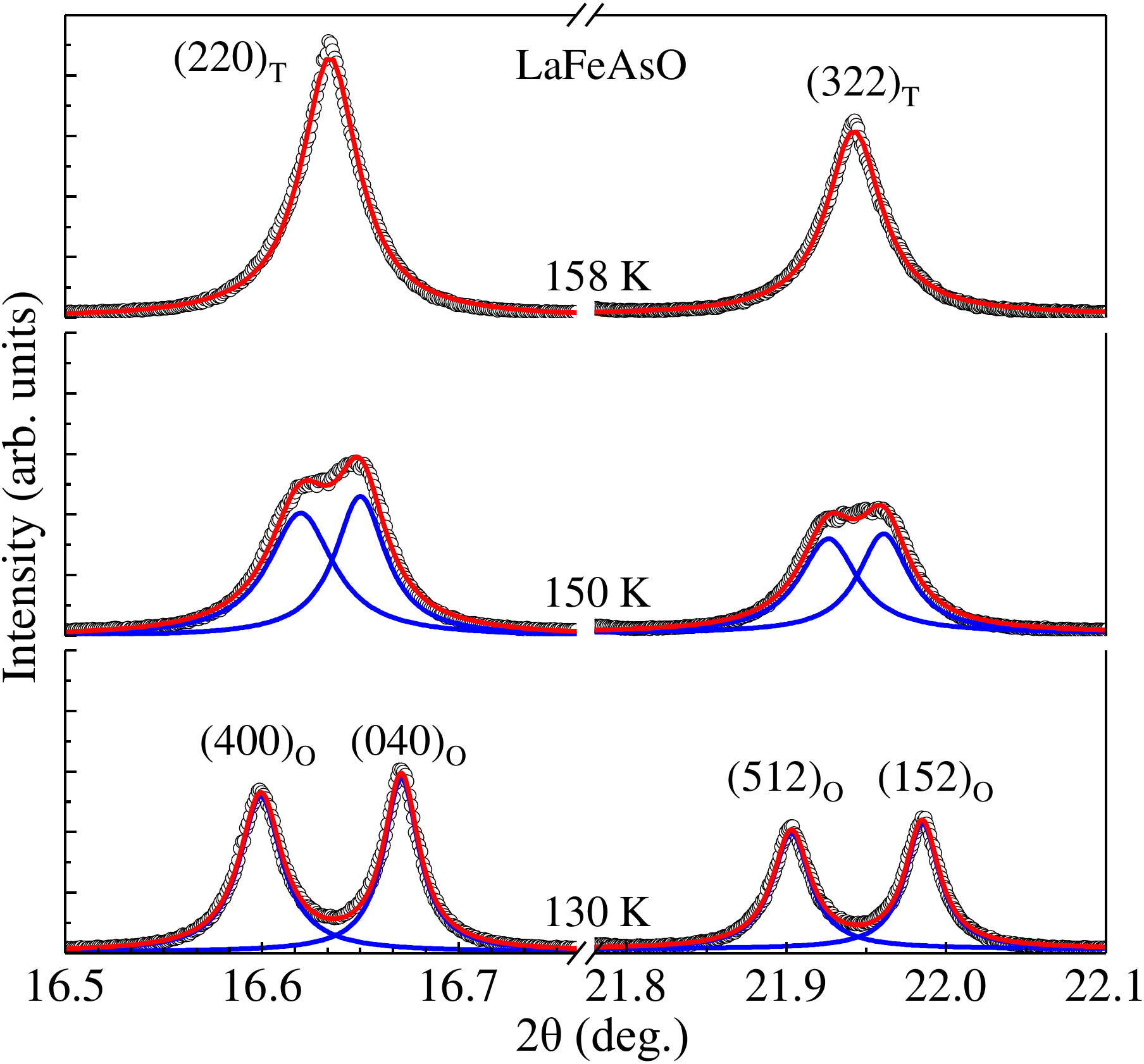}
\caption{(color online) Synchrotron X-ray powder diffraction data of pulverized LaFeAsO single crystals above (158 K), during (150 K) and below (130 K) the O-T
transition (circles). The solid lines are fits of the Lorentzian line-shape.}
\label{sy-1}
\end{figure}
\begin{figure}
\centering
\includegraphics[width = 0.46\textwidth] {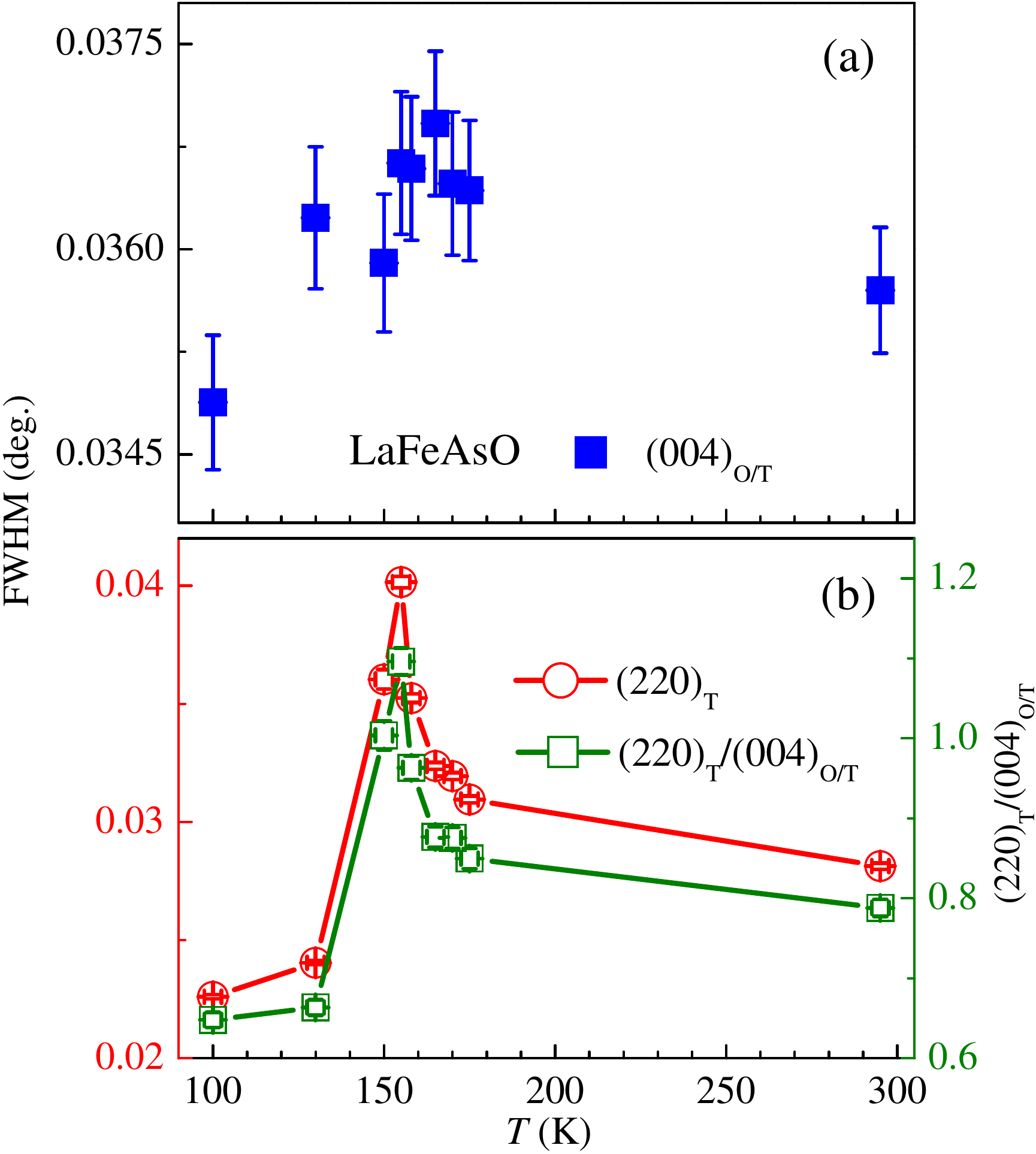}
\caption{(color online) Temperature variation of FWHM of the (004)$_{\texttt{O/T}}$ (a) and (220)$_\texttt{T}$ (circles) (b) reflections from synchrotron X-ray
powder diffraction study of pulverized LaFeAsO single crystals. Below the O-T structural transition, the FWHM of the (220)$_\texttt{T}$ reflection was averaged from
the (400)$_\texttt{O}$ and (040)$_\texttt{O}$ reflections. To avoid the influence of systematic errors, we normalized the FWHM of the (220)$_\texttt{T}$
reflection to that of the (004)$_{\texttt{O/T}}$ as shown the squares in (b). The lines in (b) are guides to the eyes.}
\label{sy-2}
\end{figure}
\begin{table}[!ht]
\caption{The \textbf{q} values, integrated intensities, and the corresponding values of $|F_\texttt{M}(hkl)|^2\sin^2\alpha$ at 9.5 K of the AFM reflections
observed in single-crystal LaFeAsO with the AFM structure as shown in Fig.\ \ref{structures}(b).} \label{table2}
\begin{ruledtabular}
\begin{tabular} {cccc}
(\emph{h}0\emph{l}) & \textbf{q} (\AA$^{-1}$) & Intensity & $|F_\texttt{M}(hkl)|^2\sin^2\alpha$ \\
\hline
(101) & 1.162   & 26.55$\pm$0.35      & 6.138   \\
(103) & 1.545   & 107.50$\pm$1.00     & 31.251   \\
(105) & 2.111   & 94.08$\pm$0.81      & 46.470   \\
(107) & 2.751   & 53.87$\pm$0.48      & 53.670   \\
(109) & 3.422   & 22.21$\pm$0.31      & 57.325   \\
(303) & 3.487   & 2.98$\pm$0.11       & 6.138   \\
(305) & 3.773   & 3.35$\pm$0.11       & 14.560   \\
(1011) & 4.110  & 7.29$\pm$0.13       & 59.373   \\
(307)  & 4.164  & 2.42$\pm$0.08       & 23.424   \\
\end{tabular}
\end{ruledtabular}
\end{table}
\begin{table}[!ht]
\caption{The fitted results of Fig.\ \ref{cnq} with different form factors $\langle j_0\rangle$ of Fe, Fe$^{2+}$ and Fe$^{3+}$ with spin contribution only
\cite{Brown1992}, and a homogeneous spherical model (details in text): the corresponding magnetic moment size \emph{M} $(\mu_\texttt{B})$ and ionic radius
\emph{R} ({\AA}). The effective ionic radius of Fe$^{2+}$ and Fe$^{3+}$ was taken from \cite{Shannon1976} with the CN = 4.} \label{table3}
\begin{ruledtabular}
\begin{tabular} {ccc}
Models & \emph{M} $(\mu_\texttt{B})$ & \emph{R} ({\AA}) \\
\hline
SrFe$_2$As$_2$/bcc-iron   &   0.83$\pm$0.04    &                       \\
Fe$^{2+}$                 &   0.82$\pm$0.03    &     0.63              \\
Fe$^{3+}$                 &   0.77$\pm$0.03    &     0.49              \\
sphere                    &   0.74$\pm$0.04    &     0.63$\pm$0.04     \\
\end{tabular}
\end{ruledtabular}
\end{table}
\begin{figure}
\centering
\includegraphics[width = 0.45\textwidth] {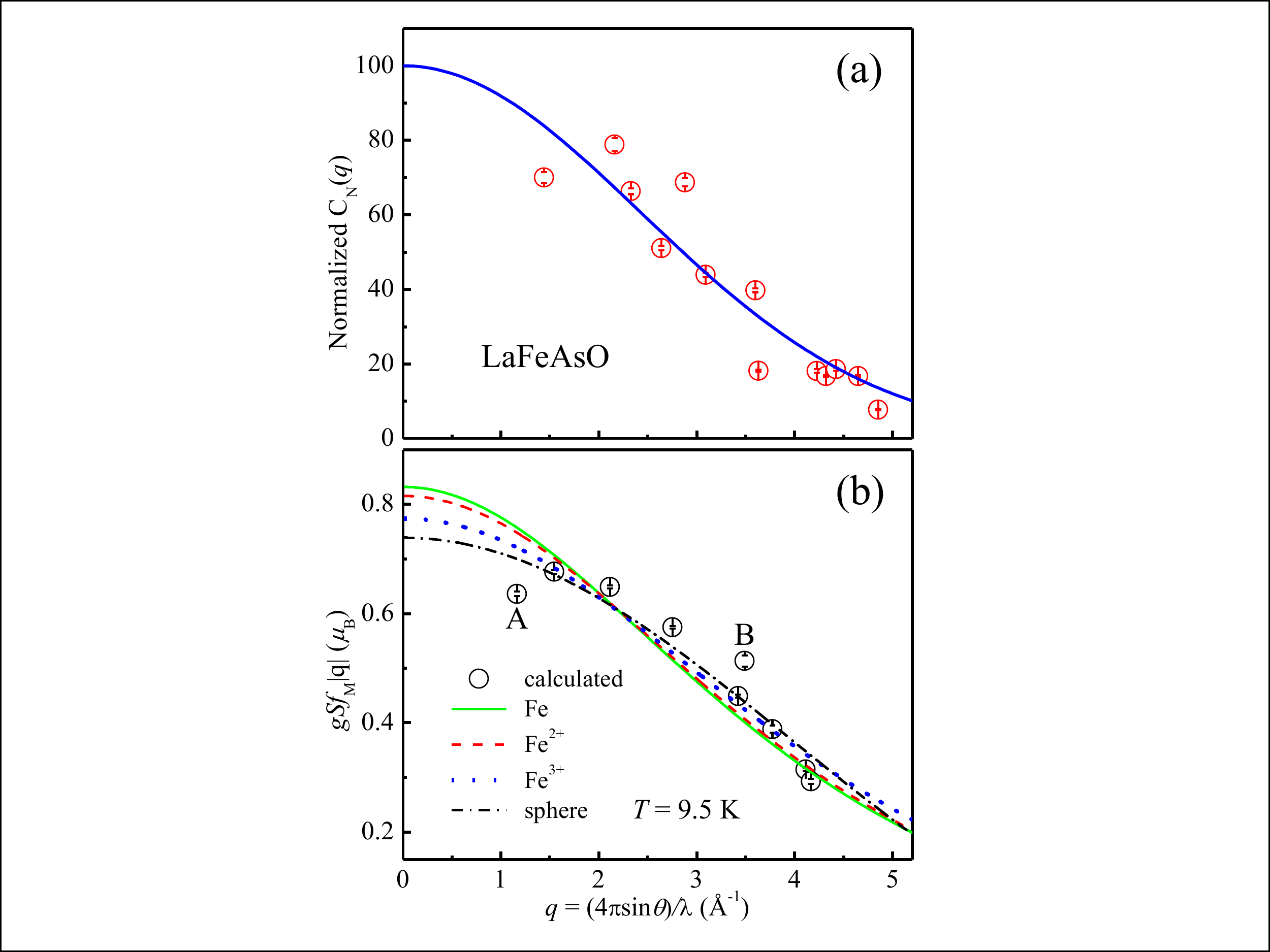}
\caption{(color online) (a) \emph{q}-dependence of the collected factor $C_\texttt{N}(q)$ (circles) in Eq. (2) at 9.5 K as described in the text. The nuclear
Bragg reflections were chosen to cover the \emph{q}-range of the measured AFM reflections as possible. The intensities
of nuclear reflections were integrated from their respective rocking curves, taking into account the domain effect, and normalized to the corresponding structure
factors and the Lorentz factors. The collected factor $C_\texttt{N}(q)$ contains the DWF and other effects such as extinctions and absorption. The solid line is
the best DWF-like curve that best fits the data and is used to normalize the AFM peaks to obtain the magnetic form factor of iron in single-crystal LaFeAsO.
(b) Measured magnetic form factor of iron (circles) at the AFM phase (9.5 K) of single-crystal LaFeAsO. The lines as indicated are the best fits of the
data with the magnetic moment \emph{M} ($\mu_\texttt{B}$) multiplied by the calculated form factors $\langle j_0\rangle$ of Fe, Fe$^{2+}$ and
Fe$^{3+}$ with spin contribution only \cite{Brown1992}, and the form factor of a homogeneous sphere (details in text), respectively. The fitted results are shown
in Tab.\ \ref{table3}. The deviation of A and B points from the smooth curves is discussed in the text.}
\label{cnq}
\end{figure}
A room-temperature X-ray powder diffraction pattern of pulverized LaFeAsO single crystals from the same batch as the one used in this study measured on an
in-house diffractometer is shown in Fig.\ \ref{powder-diff} (dots), where the best structure refinement with $P4/nmm$ symmetry using Fullprof suite
\cite{Rodriguez-Carvajal1993} is also displayed (solid line), yielding a good fit ($\chi^2$ = 1.18) and indicating a high degree of phase purity of the single
crystals. The refined structural parameters listed in Tab.\ \ref{table0} are in agreement with previous reports \cite{Yan2009,Nomura2008}.

Figure\ \ref{order-params} shows the temperature evolution of the structural and magnetic transitions in the single crystal measured by neutron diffraction
on HB-1A. For the structural transition the (400)$_\texttt{O}$/(040)$_\texttt{O}$ and (006)$_\texttt{O}$ ($Cmma$) reflections were monitored. The
integrated intensity of the rocking-curve scans of the (400)$_\texttt{O}$ reflection shows an appreciable increase at $\sim$156 K, with a relatively sharp
maximum at $\sim$148 K. The scattering at the (400)$_\texttt{O}$/(040)$_\texttt{O}$ rocking curves or $\theta$-2$\theta$ scans is strongly influenced by
twinning formation upon cooling \cite{Loudon2010} that may increase the in-plane mosaicity in the crystal due to nucleation of domains with different
orientations at different sites. Based on these observations, we suggest two critical temperatures for the structural transition: first, at
$T_{\texttt{S}}$ $\sim$ 156 K, the finite local O domains begin to form side by side with major T domains in a slightly disordered manner; second, at
$T_\texttt{P}$ $\sim$ 148 K, the local O precursors have grown into long-range O domains. The temperature range between $T_{\texttt{S}}$ and $T_\texttt{P}$
may be considered as a coexisting regime of the T and O phases. The splitting of the (\emph{HKL})$_\texttt{T}$ ($P4/nmm$) reflection into twinned $(H+K,
H-K, L)/(H-K, H+K, L) (Cmma)$ reflections is a sensitive measure of the T-to-O structural transition. To track the O splitting, we performed a
high-resolution synchrotron X-ray powder diffraction study with pulverized LaFeAsO single crystals. Part of the data are displayed in Fig.\ \ref{sy-1}
which clearly shows the splitting of (220)$_\texttt{T}$ to (400)$_\texttt{O}$/(040)$_\texttt{O}$ and (322)$_\texttt{T}$ to
(512)$_\texttt{O}$/(152)$_\texttt{O}$ at 130 K below the O-T transition. It is noted that the (220)$_\texttt{T}$ reflection at 158 K is broader than the
(400)$_\texttt{O}$ reflection at 130 K.  This may indicate a distribution of d-spacings due to O-T fluctuations above $T_\texttt{S}$. In Fig.\ \ref{sy-2},
we show the temperature variation of the FWHM of the (004)$_{\texttt{O/T}}$ and (220)$_\texttt{T}$ reflections obtained from a Lorentzian fit function to
the Bragg peaks shown in Fig.\ \ref{sy-1}. While the width variation of the (004)$_{\texttt{O/T}}$ is negligible within errors, the variation of
(220)$_\texttt{T}$ is more significant showing a maximum at $\sim$150 K.  The normalized FWHM of the (220)$_\texttt{T}$ reflection to that of the
(004)$_{\texttt{O/T}}$ reflection displays an asymmetric temperature variation with a larger FWHM above the O-T transition [Fig.\ \ref{sy-2}(b)]. This
indicates remnant orthorhombicity at temperatures as high as 300K.  This is consistent with Ref. \cite{Qureshi2010} where a finite orthorhombicity remains
visible above $T_\texttt{P}$ up to 200 K. Similar observations have also been found in SrFe$_2$As$_2$ \cite{Li2009-1,Loudon2010}. We thus argue that both T
and O phases may coexist dynamically due to the strong magnetic correlations in a certain temperature range \cite {Qureshi2010} above
\emph{T}$_\texttt{N}$, which can be ascribed to a spin nematic phase \cite{Fang2008,Jesche2010,Li2010}. We, therefore, associate $T_\texttt{P}$ = 148(1) K
as the O-T phase transition temperature in the single-crystal LaFeAsO. We argue that this transition temperature ($T_\texttt{P}$) may vary slightly,
depending on the sample quality \cite{Jesche2010}, size and shape, and the cooling/warming protocols.

Figures\ \ref{order-params}(b) and (d) show the temperature dependence of the rocking-curve integrated intensity of the AFM (103) reflection. The magnetic
peak appears at $\sim$140 K, indicating the formation of the AFM structure as shown in Fig.\ \ref{structures}(b).

\subsection{B. The magnetic form factor of iron}

To obtain the magnetic form factor of iron in LaFeAsO we collected the integrated intensities of the rocking curves of nuclear and magnetic Bragg
reflections to the highest order possible with the HB-1A spectrometer. The nuclear reflections, necessary to obtain the magnetic moment size, were chosen
to be able to cover as much of the \emph{q}-range of the measured AFM reflections as possible. This reduces any errors due to geometry corrections,
Debye-Waller factor (DWF), absorption effects, and others. The integrated intensity of the rocking curve of a nuclear Bragg reflection at a reciprocal
lattice vector \textbf{q} in a crystal is given by:
\begin{eqnarray}
I_\texttt{N} & = & \frac{V}{v_0^2}\Phi_0(\theta)|F_\texttt{N}(\textbf{q})|^2\frac{\lambda^3}{2\mu\sin (2\theta)}\mbox{ e}^{-2W} \nonumber\\
& = & \frac{C_\texttt{N}(\textbf{q})|F_\texttt{N}(\textbf{q})|^2}{\sin (2\theta)},
\end{eqnarray}
where $V$ is the scattering volume of the crystal, $v_0$ is the unit-cell volume, $\Phi_0(\theta)$ is the beam flux at the angle $\theta$,
$F_\texttt{N}(\textbf{q})$ is the structure factor, $\lambda$ is the beam wavelength, $\mu$ is the absorption length, $\sin(2\theta)$ is the Lorentz factor
for a rotating crystal, and $2W = \textbf{q}^2\langle u_\texttt{Q}\rangle^2$ is the DWF. We collect all the constants and unknown \textbf{q}-dependent
factors in $C_\texttt{N}(q)$, including the DWF. The integrated intensities of the rocking curves of the chosen nuclear Bragg reflections were normalized
to the corresponding structure factors and the Lorentz factors. The data were also corrected for the fact that the rocking curve of nuclear Bragg peaks in
our measurements includes an equivalent twinned (\emph{h}00) and (0\emph{k}0) domains \cite{Xiao2009}, for normalization of the magnetic reflections which
are due to the (\emph{h}00) domain only. The results are shown in Fig.\ \ref{cnq}(a) (circles). We assume $C_\texttt{N}(q)$ is a DWF-like function (namely,
a gaussian) and fit the data by the nonlinear square technique to obtain a smooth $C_\texttt{N}(q)$ function shown as a solid line in Fig.\ \ref{cnq}(a).

Similarly, the scattered intensity of the rocking curve of a magnetic Bragg reflection can be expressed as:
\begin{eqnarray}
I_\texttt{M} = C_\texttt{M}(\textbf{q}) (\gamma_\texttt{n} r_\texttt{e}\frac{1}{2}gS)^2f^2_\texttt{M}(|\textbf{q}|)|F_\texttt{M}(hkl)|^2
\sin^2\alpha\frac{1}{\sin (2\theta)},
\label{eqn-cmq}
\end{eqnarray}
where $C_\texttt{M}(\textbf{q})$ = $C_\texttt{N}(\textbf{q})/4$ due to the fact that the AFM unit cell is doubled along the crystallographic \emph{c} axis
in comparison with the nuclear one [Fig.\ \ref{structures}(b)], $\gamma_\texttt{n}$ = -1.913, $r_\texttt{e} = 2.81794 \times 10^{-5}$ {\AA} is the
classical electron radius, $f_\texttt{M}(|\textbf{q}|)$ is the magnetic form factor at the magnetic reciprocal lattice (\textbf{q}),
$|$$F_\texttt{M}(hkl)$$|$ = $|$$\sum e^{2 \pi i(hx_j + ky_j + lz_j)}$$|$ = 8 (\emph{j} = 1 to 8) where ($x_j$ $y_j$ $z_j$) represents fractional
coordinates of the \emph{j}th atom in the AFM unit cell, and $\sin^2\alpha=1-(\hat{\bf{q}}\cdot \hat{\bf{\mu}})^2$ where $\hat{\bf{q}}$ and
$\hat{\bf{\mu}}$ are the unit vectors along the scattering vector and the direction of the moment, respectively. Our attempt to calculate the magnetic form
factor with various directions of the ordered magnetic moment yielded an irregular and unphysical form factor, except when the moment points along the O
$a$ axis, namely, $\hat{\bf{\mu}}=$ (100). This is strong confirmation that the AFM moment direction is along the O \emph{a} axis. In Tab.\ \ref{table2} we
list all the observed magnetic reflections, their integrated intensities, and the corresponding values of $|F_\texttt{M}(hkl)|^2\sin^2\alpha$. Using these
parameters and Eq.\ (\ref{eqn-cmq}) we calculated the value of $gSf_\texttt{M}(|\textbf{q}|)$ as shown in Fig.\ \ref{cnq}(b) (circles). To obtain the
behavior of the form factor we fit the data in Fig.\ \ref{cnq}(b) to different model form factors of iron in different environments. In general, we find
that the form factor behavior is very similar to the one recently measured and calculated for SrFe$_2$As$_2$ \cite{Lee2010}.  Fitting the data with the
theoretical form factors of Fe, Fe$^{2+}$ and Fe$^{3+}$ ($\langle j_0\rangle$, spin contribution only) in Ref.\ \cite{Brown1992}, we obtain the
corresponding average AFM moment sizes \emph{M} ($\mu_\texttt{B}$) as listed in Tab.\ \ref{table3}. The experimental points indicated as A and B in Fig.\
\ref{cnq}(b) deviate from the smooth curves, which may be due to systematic errors or due to some subtle features of the magnetic structure and its
dynamics that are not captured by our structure factor determination.

As a simple, heuristic model, we assume that the magnetic moment is homogeneously distributed within a sphere of radius $R$, for which the form factor can
be simply calculated by $f(q)=3\frac{\sin(qR)-qR\cos(qR)}{(qR)^3}$. Fitting our measured data [circles in Fig.\ \ref{cnq}(b)] with this equation yields an
average moment size of (0.74$\pm$0.04) $\mu_\texttt{B}$ with a radius \emph{R} = (0.63$\pm$0.04) {\AA}. This radius is almost the same as the effective
ionic radius of Fe$^{2+}$ (0.63 {\AA}) \cite{Shannon1976} with a coordination number of 4, suggesting that the oxidation state of iron in LaFeAsO is
closer to that of Fe$^{2+}$. It is pointed out that the effective ionic radius depends on the particular electron configuration as well as the surrounding
ions (As) and the relative amount of ionic bonding.

The average magnetic moment is $\sim$0.8 $\mu_\texttt{B}$ in the AFM phase of single-crystal LaFeAsO in disagreement with the reported smaller value of
$\sim$0.37 $\mu_\texttt{B}$ (polycrystal) \cite{Cruz2008}, but more or less near the recently reported $\sim$0.63(1) $\mu_\texttt{B}$ (polycrystal)
\cite{Qureshi2010}. In addition, the obtained moment size in this study is almost the same as the first-principles-optimized 0.87 $\mu_\texttt{B}$ in the
magnetically frustrated state \cite{Yildirim2008}, the $\sim$0.83 $\mu_\texttt{B}$ in an oxygen deficient LaFeAsO (polycrystal) \cite{Nowik2008}, and the
$\sim$0.8 $\mu_\texttt{B}$ in CeFeAsO (polycrystal) \cite{zhao2008}. First, it is pointed out that there is a large difference in the properties of
polycrystalline and single-crystalline samples, especially for those containing easily-sublimating elements or oxides, e.g., the nominal
La$_{0.875}$Sr$_{0.125}$MnO$_3$ \cite{Li2007-1,Li2007-2,Li2009}. Second, single crystals are believed to be more stoichiometric \cite{Li2007-1,Li2009} and
maintain translational symmetry over macroscopic distances, in contrast with polycrystals, thereby providing more reliable information on the structures
and intrinsic properties. The small magnetic moment size $\sim$0.8 $\mu_\texttt{B}$ (compared to Fe$^{2+}$ in an insulator, $S=2$ with an average ordered
localized moment 4 $\mu_\texttt{B}$) is comparable to that of the \emph{A}Fe$_2$As$_2$ (\emph{A} = Ca, Sr, Ba, $\texttt{"}$122$\texttt{"}$) suggesting an
effective $S \approx \frac{1}{2}$ on the iron sites \cite{Lee2010}, and to some extent points to the itinerant character of the magnetism in this system.
This puts the $\texttt{"}$1111$\texttt{"}$ compound on a similar par with the other $\texttt{"}$122$\texttt{"}$ parent compounds with a magnetic moment
that is in the range of 0.8-1.1 $\mu_\texttt{B}$. This suggests a sharp jump in the magnetic moment of Fe ions as a function of the Fe-As distance to about
1 $\mu_\texttt{B}$, as predicted from first-principle calculations of bulk zinc-blende FeAs \cite{Mirbt2003}.

There are a few possible explanations to the small magnetic moment measured by neutron diffraction techniques in iron-pnictides. The most accepted one is
that it is due to the itinerant character of magnetism in this system. On the other hand, in a local moment picture this may be due to the dimensionality
of the system or due to competing nearest neighbors (NN) and next-NN (NNN) exchange interactions. The Fe layers are weakly coupled making each layer behave as a
quasi-2D system. The 2D systems, especially those with isotropic NN coupling, exhibit strong magnetic fluctuations that can lower the quasi-static ordered
magnetic moment. Second, the unusual spin arrangement observed in iron-pnictides indicates the presence of strong competing and conflicting interactions
between NN ($J_1$) and NNN ($J_2$) that can lead to strong magnetic fluctuations and a reduced static moment.

To summarize, employing both neutron and synchrotron diffraction techniques to explore the details of the structural and magnetic properties of
single-crystal LaFeAsO, we found: (1) The O-T structural transition occurs at $T_\texttt{P} \approx$ 148 K, but the finite local O precursors appear to
form already at $T_\texttt{S} \approx 156$ K. We argue that the T and O phases may coexist in the temperature range of $T_\texttt{S}$ and $T_\texttt{P}$,
and at $T_\texttt{P}$ the long-range O phase has formed. (2) The AFM structure forms at $T_\texttt{N} \approx 140$ K upon cooling with the iron moment
direction along the crystallographic \emph{a} axis in the O structural regime. (3) The average AFM moment size is comparable with that of the
\emph{A}Fe$_2$As$_2$ (\emph{A} = Ca, Sr, Ba), e.g., \emph{M} = (0.82$\pm$0.03) $\mu_\texttt{B}$ with a form factor of Fe$^{2+}$ with spin
contribution only. This moment size is significantly larger than the previously reported values. More detailed studies of the form-factor will be required
to determine the possibility of the recently predicted spin spatial anisotropy \cite{Lee2010}. This study shows that to a large extent the properties of
LaFeAsO are very similar to those of the so-called $\texttt{"}$122$\texttt{"}$ system.

\section{ACKNOWLEDGMENTS}

D. Vaknin wishes to thank R. J. McQueeney for the many illuminating discussions on the properties of FeAs-based compounds. J.-Q. Yan, R. W. McCallum and T.
A. Lograsso thank B. Jensen and K. W. Dennis for their help in crystal growth and characterization. The authors thank Matthew Suchomel for running the
powder diffraction measurements on 11-BM on the X-ray Operations and Research Beamline 11-BM at the Advanced Photon Source, Argonne National Laboratory.
Research at Ames Laboratory is supported by the U.S. Department of Energy, Office of Basic Energy Sciences, Division of Materials Sciences and Engineering
under Contract No. DE-AC02-07CH11358. The Research at Oak Ridge National Laboratory's High Flux Isotope Reactor is sponsored by the Scientific User
Facilities Division, Office of Basic Energy Sciences, U.S. Department of Energy. Use of the Advanced Photon Source at Argonne National Laboratory was
supported by the U.S. Department of Energy, Office of Science, Office of Basic Energy Sciences, under Contract No. DE-AC02-06CH11357.

\end{document}